# Automatic Generation of Topic Labels

Areej Alokaili*
areej.okaili@sheffield.ac.uk
University of Sheffield

Nikolaos Aletras
n.aletras@sheffield.ac.uk
University of Sheffield

Mark Stevenson
mark.stevenson@sheffield.ac.uk
University of Sheffield

## ABSTRACT
Topic modelling is a popular unsupervised method for identifying the underlying themes in document collections that has many applications in information retrieval. A topic is usually represented by a list of terms ranked by their probability but, since these can be difficult to interpret, various approaches have been developed to assign descriptive labels to topics. Previous work on the automatic assignment of labels to topics has relied on a two-stage approach: (1) candidate labels are retrieved from a large pool (e.g. Wikipedia article titles); and then (2) re-ranked based on their semantic similarity to the topic terms. However, these extractive approaches can only assign candidate labels from a restricted set that may not include any suitable ones. This paper proposes using a sequence-to-sequence neural-based approach to generate labels that does not suffer from this limitation. The model is trained over a new large synthetic dataset created using distant supervision. The method is evaluated by comparing the labels it generates to ones rated by humans.

## 1 INTRODUCTION
Probabilistic topic models, such as Latent Dirichlet Allocation (LDA) [10], are a family of statistical methods that uncover the latent themes in collections of documents. They have a range of applications in information retrieval, including supporting collection exploration [1, 2, 12, 23] and query expansion [26].

Topic models typically represent a document as a multinomial distribution over topics where each topic is a multinomial distribution over words. A common way of representing a topic is to list the top N terms with the highest marginal probabilities. This representation is often sufficient when the output of the topic model is used as input to another task, such as query expansion or word sense disambiguation, but may not be when the model output is presented to a user, such as within an exploratory search system. Consequently, researchers have explored a range of techniques to improve topic model output including computing topic coherence [4, 20, 21], determining optimal topic cardinality [16], corpus pre-processing [22] and topic post-processing [7].

A popular approach has been to associate alternative labels with topics since these have been shown to reduce the cognitive load required to interpret them [2, 3, 6]. For example, the topic {*pain, disorder, symptom, depression, anxiety, patient, chronic, depressive, study, psychiatric*} may be more easily interpreted if it was labelled with {*mental disorder*} . Previous work on topic labelling has mainly followed an approach that first retrieves candidate labels from a reference dataset (e.g. Wikipedia article titles) and then ranks these candidates to identify the most suitable label, e.g. [3, 6, 9, 15, 19]. Knowledge bases have also been used to label topics by matching topic words to concepts [14, 18]. Lau et al. [17] labelled topics by selecting a word from among the top N terms as the label. Others have used images as labels [3, 5]. Techniques from summarisation have also been used to create labels for topics. Cano Basave et al. [11] proposed the first such approach to label topics created from Twitter, whereas Wan and Wang [25] extracted summary sentences from a documents related to topics.

A limitation of these extractive approaches to label generation is that they are restricted to assigning labels that are found within the set of candidates. This paper presents an alternative approach that does not suffer from this limitation. It describes a neural-based model that automatically generates labels for topics in a single step, instead of retrieving and ranking candidates. This paper's contributions are to propose a new approach to the generation of textual labels for topics using neural networks and describe the creation of a synthetic dataset from Wikipedia that can be used to train the labelling model.[1]

## 2 SEQUENCE-TO-SEQUENCE TOPIC LABELLER
Our approach is based on a sequence-to-sequence model (*seq2seq*) [13, 24] that takes a sequence of terms as input and generates another sequence of terms to be used as label. Seq2seq models consist of two recurrent neural networks (RNN), one of which acts as an encoder and the other as a decoder. In general, the encoder takes as input a sequence of values $X = (x_1, \ldots, x_T)$ and transforms them into hidden representations $H = (h_1, \ldots, h_T)$ which are passed to the decoder. The decoder generates the output one symbol at a time with each symbol generated being conditioned by the hidden state and the symbols generated previously, i.e. symbol $y_t$ is predicted as $P(y_t | \{y_1, \cdots, y_{t-1}\}, X)$.

In our approach the encoder takes the topic terms as input and passes them to an embedding layer that maps them into 300 dimensional embeddings followed by a bidirectional GRU consisting of

---

[1]The data and code for the approaches described in this paper are available at https://github.com/areejokaili/topic_labelling

200 units. The forward GRU reads the input in its original order $(x_1, \ldots, x_T)$, whereas the backward GRU reads in the reverse order $(x_T, \ldots, x_1)$, thereby encoding information from the proceeding and following words. The GRU's forward output, $hf_t$, and backward output, $hb_t$, are concatenated giving the hidden state $h_t$ of $x_t$.

$$\begin{aligned} hf_t &= GRU(x_t, h_{t-1}) \\ hb_t &= GRU(x_t, h_{t-1}) \\ h_t &= [hf_t; hb_t] \end{aligned} \quad (1)$$

During decoding, labels are predicted word by word. At timestep $t$, the decoder computes the hidden state $s_t$ as follows

$$s_t = GRU(y_{t-1}, s_{t-1}, c_t) \quad (2)$$

where $y_{t-1}$ is the previous prediction that gets fed back to predict the next word and $s_{t-1}$ is the previous hidden state. Notice here that $c_t$ is a context vector computed for each target word. This approach is different from traditional encoder-decoder architectures were the last hidden state of the encoder is used to compute $C$, a context vector which is used by the decoder at every time step. The context vector $c_t$ is computed as the weighted sum over all encoder hidden states and weights $\alpha_t$ using an attention mechanism [8]:

$$\begin{aligned} e_{tj} &= a\left(s_{t-1}, h_j\right) \\ \alpha_{tj} &= \frac{\exp\left(e_{tj}\right)}{\sum_{k=1}^{T_x} \exp\left(e_{tk}\right)} \\ c_t &= \sum_{j=1}^{T_x} \alpha_{tj} h_j \end{aligned} \quad (3)$$

where $a$ is a feedforward neural network learned with the rest of the model. The weights $\alpha_t$ sum to 1 and give higher weight to a specific state, which allows the decoder to focus on this state among others.

$s_t$ is used to generate the output probability over all possible vocabulary items for the labels by passing it to a dense layer with a softmax activation function:

$$P\left(y_t | \{y_1, \cdots, y_{t-1}\} X\right) = Dense(s_t) \quad (4)$$

Finally, the probability distribution resulting from eq. (4) is used to choose the word with the highest probability as the prediction $y_t$

$$y_t = argmax(P\left(y_t | \{y_1, \cdots, y_{t-1}\}, X\right)) \quad (5)$$

Figure 1 shows the neural network architecture described above when topic terms are passed as the input and a label is predicted.

## 3 DATA
### 3.1 Training Data
A set of topics represented by lists of terms and their associated labels is required to train our model. However, the current available datasets are too small to train large neural networks (e.g. Bhatia et al. [9] released a dataset that contains 228 topics with 19 labels for each). Therefore, a distant supervision approach was applied to create two different datasets consisting of pairs of topics and labels:

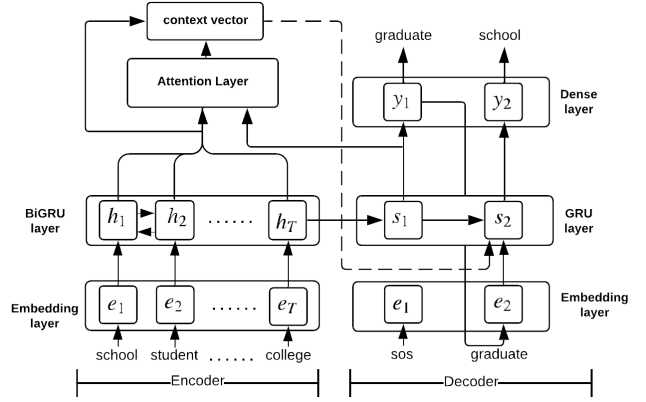

Figure 1: Illustration of the model used to label topics given the topic top n words. Diagram shows timestep $t = 2$ during which the second label word is predicted.

- **ds_wiki_tfidf** was created by selecting pairs of titles and articles from Wikipedia[2]. The article titles are treated as the *labels*, and the top 30 words from each article ranked by TFIDF are treated as synthetic *topic terms*.
- **ds_wiki_sent** is a variation of **ds_wiki_tfidf**. Rather than extracting the top 30 words using TFIDF, the *first* 30 words from the article were used as *topic terms*.

Using this approach, we collected 300,000 pairs of topics and labels for each dataset and divided into train, validate, and test sets consisting of 226,282, 12,424 and 11,800 pairs, respectively. Standard preprocessing steps were applied to clean and tokenize the datasets including the removal of numbers, special characters, rare terms and stop words[3]. The article titles (i.e. labels) in both datasets contain 13,947 unique words while the lists of topic terms contain 181,793 in **ds_wiki_tfidf** and 87,446 in **ds_wiki_sent**. Samples of both datasets are shown in Table 1.

### 3.2 Test Data
Labels generated by the model were evaluated by comparing them against gold-standard labels from two datasets. The first, described by Bhatia et al. [9] (**topics_bhatia**), contains 228 topics from four different domains (blogs, books, news and PubMed) generated by Lau et al. [15].

Bhatia et al. [9] associated each topic with 19 candidate labels by matching the topic's top 10 terms with Wikipedia titles using neural embeddings. Human ratings for those candidate labels were collected by formulating a crowdsourcing task on Amazon Mechanical Turk (MTurk). Annotators (i.e. crowdworkers) gave ratings for the labels between 0 and 3, where 3 is the highest rating. Only labels that received a high average rating (of 2 or above) were used for the dataset, resulting in 219 topics and 1156 pairs (instead of 4332, i.e. 228 topics × 19 labels).

The second dataset, **topics_bhatia_tfidf**, is an extended version of **topics_bhatia** that includes 20 additional terms for each topic.

---
[2]Using the dump enwiki-2019201-pages-articles1.xml-p10p30302
[3]Stop words were not removed from headlines.

Table 1: Samples of topics and labels from the datasets described in Section 3. Additional terms added to the topic are shown underlined.

| | Topic Terms/Article | Label/Title |
|---|---|---|
| **topics_bhatia** | oil energy gas water power fuel global price plant natural | biofuel |
| **topics_bhatia_tfidf** | oil energy gas water power fuel global price plant natural <u>lng regasification plants cold gasification turbine exhaust viable floating fluid usage conventional temperature joule acceptability argon utilisation byproducts urea cryogenic</u> | biofuel |
| **ds_wiki_tfidf** | uruguay uruguayan immigration spaniards immigrants amerindians european th argentina backbone italians background society syrian fructuoso countries matanza paraguayans bolivians uruguayans peruvians venezuelans americans colonial multiethnic del amerindian brazil people | immigration to uruguay |
| **ds_wiki_sent** | immigration uruguay started arrival spanish settlers colonial period known banda oriental immigration uruguay similar towards immigration argentina throughout history uruguay known gain massive waves immigration around world specifically european immigration | immigration to uruguay |

These additional terms were added to the 10 from **topics_bhatia** so that each topic consists of 30 terms, matching the encoder length. Additional terms were identified by finding documents associated with each topic and choosing the 20 with the highest TFIDF scores. Unfortunately the topic-document distributions are not available for **topics_bhatia**. Consequently suitable documents were identified by computing cosine similarity between the topic terms and documents using word embeddings. While the lack of information about the topic-document distributions is far from ideal, we chose to use **topics_bhatia** since it provides ratings for labels and these are expensive to obtain. Samples are shown in Table 1.

## 4 EXPERIMENTAL SETUP

### 4.1 Hyperparamters

Model hyperparameters were tuned by randomly sampling more than 50 combinations and choosing the one that produced the smallest loss on the validation set. Combinations are drawn from: optimizer [adam, rmsprop], number of encoder BiGRU layers [1,2], number of decoder GRU layers [1,2], GRU size [50, 100, 200, 300, 400, 500], dropout [0.1, 0.2,0.3, 0.4], learning rates [1e-2, 1e-3, 1e-4, 1e-5], and embedding dimensions [200, 300, 400]. As a result, the following hyperparamters were selected: Adam with learning rate 0.001 and sparse categorical cross entropy loss, one BiGRU layer for the encoder (with 200 hidden units), one GRU layer in the decoder (with 200 hidden units) and dropout of 0.1. The embedding layer was set to 300 latent dimensions and learned from scratch.[4]

### 4.2 Baselines

The labels generated by our models were compared with two baselines: the top two terms, in terms of highest marginal probabilities, for a topic (**Top-2 label**) and top three terms (**Top-3 label**).

### 4.3 Label Evaluation

BERTScore [27] was used to evaluate the quality of the generated labels.[5] BERTScore is a measure that computes the similarity between predictions and references using contextual embeddings that has shown to have high correlation with human judgments. Since BERTScore does not rely on exact matches between predicted and gold-standard labels, it is able to identify appropriate label words that do not appear in the gold labels.

Pairwise BERTScores between the topic's generated label $l$ and the gold labels $(gold\_l_1, ..., gold\_l_n)$ is computed as follows:

$$score\_topic_t = \max_{i=[1,...,n]} BERTScore(l_t, gold\_l_{ti})$$

The model's overall score is the mean score over all topics:

$$score\_model = \frac{1}{T}\sum_{t=1}^{T} score\_topic_t$$

## 5 RESULTS AND DISCUSSION

The labelling model was trained on the two Wikipedia datasets (**ds_wiki_tfidf** and **ds_wiki_sent**) where it was provided with the articles as inputs and learned to predict suitable titles. The model was used to generate labels for the Bhatia et al. [9] topics either by passing the topics' top 10 terms (**topics_bhatia**) or by passing the top 10 terms and 20 additional terms (**topics_bhatia_tfidf**). Results are shown in Table 2. All variations of our model produce significantly higher scores than the baselines of selecting the top two or three terms with the highest marginal probability as labels. The highest scores are obtained when the labelling model was trained on **ds_wiki_sent** and tested using **topics_bhatia_tfidf**. Inclusion of the 20 addition terms to the topics in **topics_bhatia_tfidf** improves results regardless of the training data used. Identifying terms using TFIDF also appears to be a useful strategy for generating training data since the highest score obtained using the **topics_bhatia** gold labels is produced by the model trained using **ds_wiki_tfidf**.

Sample labels generated using our model are shown in Table 3. The topics and gold labels are taken from **topics_bhatia**. It can be seen that the labels generated by the model are within the correct domain and similar to the gold labels. In some cases where the model did not generate an entirely new label, it picked one or two words from the topic. For example the topic {*military, force, war, army, soldier, gun, fire, air, guard, u.s*} was labeled with {*military force*} by the model trained on **ds_wiki_sent**.

An error analysis was carried out to examine cases where the model produced sub-optimal labels. For example, the topic {*mr, mrs, young, lady, look, friend, tell, mother, miss, father*} was labelled with {*the*} which may be due to the topic being incoherent and with no obvious theme.

---
[4]Pre-trained embeddings were not found to improve performance.
[5]Results were generated using the reference implementation: https://github.com/Tiiiger/bert_score

Table 2: Average BERTScore between predicted and gold labels. Each predicted label is compared to a set of gold labels to measure appropriateness as described in Section 4.3.

| | | | | BERTScore | | |
|---|---|---|---|---|---|---|
| | | | | P | R | F |
| **Baselines** | | | Top-2 label | 0.902 | 0.912 | 0.902 |
| | | | Top-3 label | 0.870 | 0.903 | 0.882 |
| Train data | ds_wiki_tfidf | Test data | topics_bhatia | 0.922*† | 0.928*† | 0.922*† |
| | | | topics_bhatia_tfidf | 0.926*† | 0.930*† | 0.925*† |
| | ds_wiki_sent | | topics_bhatia | 0.919† | 0.926† | 0.919† |
| | | | topics_bhatia_tfidf | **0.930***† | **0.933***† | **0.929***† |

* and † denote statistically significant difference ($p < 0.001$) compared to Top-2 label and Top-3 label, respectively.

Table 3: Samples of labels produced by variations of the models trained using the ds_wiki_tfidf and ds_wiki_sent datasets.

| | Model trained on | |
|---|---|---|
| | ds_wiki_tfidf | ds_wiki_sent |
| **Topic 1** | vote house election poll bill republican party voter candidate senate | |
| **Gold labels (Top 5)** | election, by-election, general election, primary election, electoral college | |
| topics_bhatia | hall of representatives elections | the house |
| topics_bhatia_tfidf | united states house of representatives elections in illinois | united states presidential election |
| **Topic 2** | plane kennedy flight fly pilot airline airport air search passenger | |
| **Gold labels (Top 5)** | airplane, boeing 737, airliner | |
| topics_bhatia | group | plane and |
| topics_bhatia_tfidf | the real flight | the vanishing of flight |
| **Topic 3** | fight lewis ray bob hoya boxing ring king vegas champion | |
| **Gold labels (Top 5)** | super middleweight, professional boxing, light middleweight | |
| topics_bhatia | lewis | fight lewis and the |
| topics_bhatia_tfidf | lewis | fight lewis and hoya |
| **Topic 4** | military force war army soldier gun fire air guard u.s. | |
| **Gold labels (Top 5)** | united states army, united states marine corps, military police, artillery, aerial warfare | |
| topics_bhatia | royal army | military force war |
| topics_bhatia_tfidf | operation combat force | military force |

The topic {*artery, vascular, coronary, stent, vein, vessel, carotid, aortic, aneurysm, arterial*} was labelled by {*and*}, which is due to the topic being from a different domain to the data used to train the model (since Wikipedia is a general domain resource).

## 6 CONCLUSION

We presented the first seq2seq model to generate textual labels for automatically generated topics. We also presented a dataset built from Wikipedia that is used to train the labelling model. BERTScore was used to measure the similarities between the generated labels and gold standard labels. We find seq2seq models to be a generic approach that produces appropriate labels. We observe that there is margin for improvements that may generate more appropriate labels.

## REFERENCES

[1] Nikolaos Aletras, Timothy Baldwin, Jey Lau, and Mark Stevenson. 2014. Representing Topics Labels for Exploring Digital Libraries. In *Proc. of JCDL '14*. London United Kingdom, 239–248.
[2] Nikolaos Aletras, Timothy Baldwin, Jey Han Lau, and Mark Stevenson. 2017. Evaluating topic representations for exploring document collections. *JASIST* 68, 1 (2017), 154–167.
[3] Nikolaos Aletras and Arpit Mittal. 2017. Labeling Topics with Images using Neural Networks. In *Proc. of ECIR '17*. Aberdeen, UK, 500–505.
[4] Nikolaos Aletras and Mark Stevenson. 2013. Evaluating Topic Coherence using Distributional Semantics. In *Proc. of IWCS '13*, Vol. 13. Potsdam, Germany, 13–22.
[5] Nikolaos Aletras and Mark Stevenson. 2013. Representing Topics Using Images. In *Proc. of NAACL HLT '13*. Atlanta, Georgia, 158–167.
[6] Nikolaos Aletras and Mark Stevenson. 2014. Labelling Topics using Unsupervised Graph-based Methods. In *Proc. of ACL '14*. Baltimore, Maryland, 631–636.
[7] Areej Alokaili, Nikolaos Aletras, and Mark Stevenson. 2019. Re-Ranking Words to Improve Interpretability of Automatically Generated Topics. In *Proc. of IWCS '19*. Gothenburg, Sweden, 43–54.
[8] Dzmitry Bahdanau, Kyung Hyun Cho, and Yoshua Bengio. 2015. Neural machine translation by jointly learning to align and translate. In *Proc. of ICLR '15*. International Conference on Learning Representations, ICLR, San Diego, CA. arXiv:1409.0473
[9] Shraey Bhatia, Jey Lau, and Timothy Baldwin. 2016. Automatic Labelling of Topics with Neural Embeddings. In *Proc. of COLING '16*. Osaka, Japan, 953–963.
[10] David Blei, Andrew Ng, and Michael Jordan. May 2003. Latent Dirichlet Allocation. *Journal of machine Learning research* 3, 1 (May 2003), 993–1022.
[11] Amparo Elizabeth Cano Basave, Yulan He, and Ruifeng Xu. 2014. Automatic Labelling of Topic Models Learned from Twitter by Summarisation. In *Proc. of ACL '14*. Baltimore, USA, 618–624.
[12] Allison Chaney and David Blei. 2012. Visualizing Topic Models. In *Proc. of ICWSM '12)*. Dublin, Ireland, 419–422.
[13] Kyunghyun Cho, Bart Van Merriënboer, Caglar Gulcehre, Dzmitry Bahdanau, Fethi Bougares, Holger Schwenk, and Yoshua Bengio. 2014. Learning phrase representations using RNN encoder-decoder for statistical machine translation. In *Proc. of EMNLP '14*. Doha, Qatar, 1724–1734.
[14] Ioana Hulpus, Conor Hayes, Marcel Karnstedt, and Derek Greene. 2013. Unsupervised graph-based topic labelling using dbpedia. In *Proc. of WSDM '13*. Rome, Italy, 465–474.
[15] Jey Lau, Karl Grieser, David Newman, and Timothy Baldwin. 2011. Automatic Labelling of Topic Models. In *Proc. of ACL HLT '11*. Portland, Oregon, 1536–1545.
[16] Jay H. Lau and Timothy Baldwin. 2016. The Sensitivity of Topic Coherence Evaluation to Topic Cardinality. In *Proc. of NAACL-HLT '16*. San Diego, CA, 483–487.
[17] Jey H. Lau, David Newman, Sarvnaz Karimi, and Timothy Baldwin. 2010. Best topic word selection for topic labelling. In *Proc. of COLING '10*. Beijing, China, 605–613.
[18] Davide Magatti, Silvia Calegari, Davide Ciucci, and Fabio Stella. 2009. Automatic Labeling of Topics. In *Proc. of ISDA '09*. Pisa, Italy, 1227–1232.
[19] Qiaozhu Mei, Xuehua Shen, and Chengxiang Zhai. 2007. Automatic Labeling of Multinomial Topic Models. In *Proc. of ACM SIGKDD '07*. San Jose, California, 490–499.
[20] David Mimno, Hanna Wallach, Edmund Talley, Miriam Leenders, and Andrew Mccallum. 2011. Optimizing Semantic Coherence in Topic Models. In *Proc. of EMNLP '11*. Edinburgh, Scotland, 262–272.
[21] David Newman, Han Lau, Karl Grieser, and Timothy Baldwin. 2010. Automatic Evaluation of Topic Coherence. In *Proc. of NAACL HLT'10*. Los Angeles, California, 100–108.
[22] Alexandra Schofield, Måns Magnusson, and David Mimno. 2017. Pulling Out the Stops: Rethinking Stopword Removal for Topic Models. In *Proc. of EACL '17*. Valencia, Spain, 432–436.
[23] Alison Smith, Tak Lee, Forough Poursabzi-Sangdeh, Jordan Boyd-Graber, Niklas Elmqvist, and Leah Findlater. 2017. Evaluating Visual Representations for Topic Understanding and Their Effects on Manually Generated Topic Labels. *Transactions of the Association for Computational Linguistics* 5 (2017), 1–15.
[24] Ilya Sutskever, Oriol Vinyals, and Quoc V Le. 2014. Sequence to sequence learning with neural networks. In *Proc. in NIPS '14*. Montreal, Canada, 3104–3112.
[25] Xiaojun Wan and Tianming Wang. 2016. Automatic Labeling of Topic Models Using Text Summaries. In *Proc. of ACL '16*. Berlin, Germany, 2297–2305.
[26] Xing Yi and James Allan. 2009. A comparative study of utilizing topic models for information retrieval. In *Proc. of ECIR '09*. Springer, Toulouse, France, 29–41.
[27] Tianyi Zhang, Varsha Kishore, Felix Wu, Kilian Q. Weinberger, and Yoav Artzi. 2019. BERTScore: Evaluating Text Generation with BERT. *ArXiv* abs/1904.09675 (2019).